\documentclass[twocolumn]{aastex631}

\DeclareUnicodeCharacter{2212}{-}

\shorttitle{Mass–Metallicity Trends in Transiting Exoplanets}
\shortauthors{Sun et al.}

\graphicspath{{./}{figures/}}

\begin{document}
		
\title{A revisit of the Mass–Metallicity Trends in Transiting Exoplanets}
		
\correspondingauthor{Qinghui Sun; Sharon Xuesong Wang}
\email{qingsun@mail.tsinghua.edu.cn; sharonw@mail.tsinghua.edu.cn}

\author[0000-0003-3281-6461]{Qinghui Sun}
		
\author[0000-0002-6937-9034]{Sharon Xuesong Wang}
\affiliation{Department of Astronomy, Tsinghua University, Beijing, 100084, China}
		
\author[0000-0003-0156-4564]{Luis Welbanks}
\affiliation{The School of Earth and Space Exploration, Arizona State University, Tempe, AZ 85287, USA}
		
\author[0009-0008-2801-5040]{Johanna Teske}
\affiliation{Earth and Planets Laboratory, Carnegie Institution for Science, 5241 Broad Branch Road, NW, Washington, DC 20015, USA}
		
\author[0000-0003-0426-6634]{Johannes Buchner}
\affiliation{Max-Planck-Institut f\"{u}r extraterrestrische Physik, Giessenbachstrasse 1, D-85748 Garching, Germany}
		
\begin{abstract}

The two prevailing planet formation scenarios, core-accretion and disk instability, predict distinct planetary mass-metallicity relations. Yet, the detection of this trend remains challenging due to inadequate data on planet atmosphere abundance and inhomogeneities in both planet and host stellar abundance measurements. Here we analyze high-resolution spectra for the host stars of 19 transiting exoplanets to derive the C, O, Na, S, and K abundances, including planetary types from cool mini-Neptunes to hot Jupiters ($T_{\rm eq}\ \sim$ 300 - 2700 K; planet radius $\sim$ 0.1 - 2 $R_{\rm J}$). Our Monte Carlo simulations suggest that the current dataset, updated based on Welbanks et al. 2019, is unable to distinguish between a linear relation and an independent distribution model for the abundance-mass correlation for water, Na, or K. To detect a trend with strong evidence (Bayes factor $>$ 10) at the 2$\sigma$ confidence interval, we recommend a minimum sample of 58 planets with HST measurements of water abundances coupled with [O/H] of the host stars, or 45 planets at the JWST precision. Coupled with future JWST or ground-based high resolution data, this well-characterized sample of planets with precise host star abundances constitute an important ensemble of planets to further probe the abundance-mass correlation.

\end{abstract}
		
\section{Introduction} \label{sec:intro}
		
Atmospheric chemical abundances of exoplanets can reveal important information about their formation and migration histories (\citealt{2022ExA....53..323H}). Core-accretion and disk instability are the two representative planetary formation models. The core-accretion scenario (e.g., \citealt{1996Icar..124...62P}) proposes that planetesimals accrete to form a heavy element core, which then grows and accretes gas from the surrounding disk, in a timescale of several megayears or more. Due to the condensation and migration of solids within the disk, the gas formed during the star's formation process may not have the same composition as the parent star (\citealt{2010fee..book..157L, 2011ApJ...743L..16O}). The outer disk formation followed by migration (e.g. \citealt{2018ARAA..56..175D}; \citealt[][hereafter W19]{2019ApJ...887L..20W}; \citealt{2022MNRAS.509..894H}) in the core accretion scenario would lead to a correlation in planetary mass and atmospheric metallicity for giant planets. Several studies found an inverse correlation between giant planet heavy element mass and planetary mass and stellar metallicity (e.g.  \citealt{2011ApJ...736L..29M}; \citealt{2013ApJ...775...80F}; \citealt{2016ApJ...832...41M}; \citealt{2016ApJ...831...64T}). Alternatively, the disk instability scenario (e.g. \citealt{1997Sci...276.1836B}) suggests that a disk can fragment into a dense core that can contract to form giant gaseous protoplanets in a few hundred years. However, planets formed due to disk instability are not expected to have the mass-metallicity correlation (e.g., \citealt{2016ApJ...829..114B, 2022AA...665L...5K}). Higher metallicity leads to higher optical depth, effectively trapping more radiation and subsequently raising temperatures within the disk. This elevated temperature can hinder the formation of a dense core. As a result, the formation of gas giant planets through disk instability is more favorable in stars with lower metallicity, and no observable mass-metallicity trend would emerge (\citealt{2002ApJ...567L.149B}).

The advancements in observational techniques and instruments have enabled numerous studies to measure planetary atmospheric abundances using transmission spectra. Thus far, H$_2$O, Na, and K are the most observed chemical species, as demonstrated by various studies (e.g., \citealt{2002ApJ...568..377C, 2008ApJ...673L..87R, 2014ApJ...791L...9M, 2015AA...577A..62W, 2017ApJ...834...50B, 2018Natur.557..526N, 2019MNRAS.482.1485P}). These species have been utilized to compare exoplanetary atmosphere compositions (e.g., \citealt{2016Natur.529...59S, 2016ApJ...826L..16H}). By normalizing abundances to the compositions of host stars, we can establish constraints on the differential enhancements of elements in the atmospheres of hot gas giants compared to their host stars. This normalization is critical for uncovering the most precise mass-metallicity trend. Prior studies emphasize the significance of a homogeneous dataset in analyzing trends using exoplanetary atmospheric data (e.g. \citealt{2023ApJS..269...31E}). Furthermore, homogeneous measurements of elemental abundance in host stars hold intrinsic significance. For instance, \citet{2022AJ....164...87K} provide chemical abundance analyses for 17 FGK stars, which serve as host stars of exoplanets observed in the Cycle-1 JWST observer programs. However, not all studies have performed an ``apples-to-apples" comparison of homogeneously-derived planet and stellar abundances, due to different instruments and data reduction methods, which could introduce scatter into any observed trend.
		
\citet[][hereafter W19]{2019ApJ...887L..20W} report Na, K, and H$2$O abundances for a selection of 19 exoplanets, by using transmission spectra gathered from the Hubble Space Telescope (HST) Wide Field Camera 3 (WFC3), Space Telescope Imaging Spectrograph (STIS), and Spitzer photometry. All the {\it HST} data used in this paper can be found in MAST: \dataset[10.17909/T97P46]{http://dx.doi.org/10.17909/T97P46}. The sample encompasses a range of exoplanets from cool mini-Neptunes to hot Jupiters, exhibiting equilibrium temperature ($T{\rm eq}$) between 300 to 2700 K and planet radii between 0.1 - 2 $R_{\rm J}$. The H$_2$O abundances were converted to oxygen under the assumption of thermochemical equilibrium, and the resulting values indicate a general depletion of oxygen. In instances where stellar Na and K abundances were absent, [Fe/H] (or [M/H]) was substituted. However, the lack of stellar Na and K abundances in the literature, coupled with systematic discrepancies among these values complicates the establishment of a robust correlation, underscoring the need for a homogeneous abundance analysis of the host stars.

In this paper, we reassess the mass-metallicity trend for H$_2$O by comparing the W19 planet atmosphere abundances to their homogeneously analyzed host abundances, which include C, O, Na, S, and K. This dataset not only provides crucial parameters for future inquiries but also aids in establishing the requisite number of planets needed to discern between distinct planet formation scenarios during the JWST era.

\section{Stellar Abundances} 
		
\subsection{Data Acquisition and Stellar Atmospheres}

Spectra for the host stars of the 19 exoplanets from \citet{2019ApJ...887L..20W} were obtained from the European Southern Observatory (ESO) archive, with reduced spectra from either the HARPS (\citealt{https://doi.org/10.18727/archive/33}) or FEROS (\citealt{https://doi.org/10.18727/archive/24}) instrument, or reduced HIRES spectra from the Keck Observatory Archive (KOA) (\citealt{Tran2014}). The spectra were corrected for radial velocity offset and co-added to achieve a higher signal-to-noise ratio (SNR), which was used for equivalent width measurements. Details on the sources of the spectra, estimated SNR, and principal investigators (PIs) of the datasets can be found in Table \ref{tab:NLTE}.

We adopt stellar parameters from the literature, including effective temperature ($T_{\rm eff}$), log {\it g}, microturbulence ($V_{\rm t}$), and metallicity ([Fe/H]). The stellar atmosphere and references are shown in Table \ref{tab:atmosphere}. The host stars span from late-A type dwarfs ($\sim$ 7500 K) to early-M dwarfs ($\sim$ 3500 K).

\begin{deluxetable*}{cccp{4cm}|ccccc}
	\label{tab:NLTE}
	\tablecaption{Source of spectra and non-LTE correction}
	\tablehead{
		\multicolumn4c{Source of Spectra} & \multicolumn5c{NLTE correction$^a$} \\
		Star name & Instrument & SNR & PI(s) &  \colhead{$\delta(C)$} &  \colhead{$\delta(O)$} &  \colhead{$\delta(Na)$} & \colhead{$\delta(S)$}& \colhead{$\delta(K)$} \\
	}
	\colnumbers
	\startdata
	K2-18 & HARPS & 30 & Bonfils, Xavier & -- & -- & 0.0 & -- & -- \\
	GJ 3470 & HARPS & 30 & Bonfils, Xavier; Pepe, Francesco & -- & -- & 0.0 & -- & -- \\
	HAT-P-26 &FEROS & 120 & Sousa, Sergio & 0.0 &	-0.09 & -0.1 & 0 & -0.15 \\
	HAT-P-11 & HIRES & 70 & Ford, Eric & 0.0 & -0.02 & -0.1	& 0	& -0.08 \\
	WASP-107 & HARPS, FEROS & 50, 60 & Ehrenreich, David; Jones, Matias; Triaud, Amaury & 0 & -0.01 & -0.1 & 0 & -0.11 \\
	WASP-127 & HARPS, FEROS & 190, 340 & Ehrenreich, David; Neveu Van Malle, Marion; Sarkis, Paula & -0.02 & -0.2 & -0.1	& -0.16	& -0.4\\ 
	HAT-P-12 & HIRES & 60 & Bakos, Gaspar & -0.01	& -0.02	& -0.1	& 0.0 &	-0.07 \\
	WASP-39 & FEROS & 90 & Sarkis, Paula & -0.01 & -0.11 & -0.1	& -0.14	& -0.03 \\
	WASP-31 & HARPS & 140 & Cameron, A.; Ehrenreich, David & -0.01 & -- & -0.1	& -0.3	& -- \\
	WASP-96 & FEROS & 60 & Sarkis, Paula & 0.0 &	-0.09	& -0.1	& -0.11	& -0.02 \\
	WASP-6 & HARPS & 40 & Cameron, A.; Ehrenreich, David; Oshagh, Mahmoudreza & -0.01	& -0.11	& -0.1	& -0.13	& -0.41 \\
	WASP-17 & HARPS, FEROS & 120, 140 & Cameron, A.; Ehrenreich, David; Mayor, Michel; Observatory, La Silla; Triaud, Amaury; Faedi, Francesca  & -0.03 & -0.34 & -0.1	& -0.06	& -0.62 \\
	HAT-P-1 & HIRES & 180 & Asplund, Martin & -0.03	& -0.24	& -0.1	& 0	& -0.33 \\
	HD 209458 & HARPS, FEROS & 130, 200 & HARPS User; Kuerster, M.; Mundt, Reinhard & -0.02	& -0.27	& -0.1	& -0.09	& -0.5 \\
	HD 189733 & HARPS, FEROS & 110, 120   & Lecavelier Des Etangs, A.; Mayor, Michel; Mundt, Reinhard & -0.01 &	-0.03 &	-0.1 &	-0.06 &	-0.14 \\
	WASP-19 & HARPS, FEROS & 30, 60    & Cameron, A.; Csizmadia, Szilard; Oshagh, Mahmoudreza; Faedi, Francesca; & -0.01 & -0.08 & -0.1 & -0.12 & -0.11 \\
	WASP-12 & FEROS & 100 & Sousa, Sergio & -0.02	& -0.18	& -0.15	& -0.2	& -0.52 \\
	WASP-43 & HARPS & 40 & Csizmadia, Szilard; Ehrenreich, David; Triaud, Amaury & -0.01	& -- & -0.1	& 0.0 &	-- \\
	WASP-33 & HIRES & 150 & Howard, Andrew & 0.0 &	-0.4 &	-0.1 & -- & --\\
	\enddata
	\tablecomments{a}{We adopted empirical non-LTE corrections for C, O, Na, S, and K from \citet{2022AJ....164...87K}. [X/H] = A(X) + $\delta(X)$ -A$_{\odot}$(X). }
\end{deluxetable*}

\begin{deluxetable*}{cccccc}
			\label{tab:atmosphere}
			\tablecaption{Stellar atmosphere}
			\tablehead{
				\multicolumn6c{Stellar atmospheric parameters} \\
				Star name & $T_{\rm eff}$ & log {\it g} & $V_{\rm t}$ & \colhead{[Fe/H]} & \colhead{References} \\
			}
			\colnumbers
			\startdata
			K2-18 & 3503 $\pm$ 60	& 5 (adopted)	& 1 & 0.123 & \text{W19}                                                   \\
			GJ 3470 & 3652 $\pm$ 50	& 4.78 $\pm$ 0.12	& 1 & 0.17 $\pm$ 0.06 & \text{\citet{2017AA...600A.138C}}              \\
			HAT-P-26 & 5079 $\pm$ 88 & 4.5 $\pm$ 0.5	& 0.8	& −0.04 $\pm$ 0.08 & \text{\citet{2011ApJ...728..138H}}        \\
			HAT-P-11 & 4780 $\pm$ 50 & 4.59 $\pm$ 0.03	& 0.8	& +0.31 $\pm$ 0.05 & \text{\citet{2010ApJ...710.1724B}}        \\
			WASP-107 & 4430 $\pm$ 120	& 4.5 $\pm$ 0.1	& 0.8 & +0.02 $\pm$ 0.10 & \text{\citet{2017AA...604A.110A}}           \\
			WASP-127 & 5750 $\pm$ 100 &	3.9 $\pm$ 0.1 & 1.83 & −0.18 $\pm$ 0.06 & \text{\citet{2017AA...599A...3L}}            \\
			HAT-P-12 & 4500 $\pm$ 250	& 4.0 $\pm$ 0.2	& 0.8	& −0.29 $\pm$ 0.05 & \text{\citet{2009ApJ...706..785H}}        \\
			WASP-39 & 5400 $\pm$ 150 &	4.4 $\pm$ 0.2 & 0.9 $\pm$ 0.2	& −0.12 $\pm$ 0.10 & \text{\citet{2011AA...531A..40F}} \\
			WASP-31 & 6300 $\pm$ 100	& 4.4 $\pm$ 0.1	& 1.4 $\pm$ 0.1	& −0.20 $\pm$ 0.09 & \text{\citet{2011AA...531A..40F}} \\
			WASP-96 & 5500 $\pm$ 150	& 4.25 $\pm$ 0.15	& 1.175	& +0.14 $\pm$ 0.19 & \text{\citet{2014MNRAS.440.1982H}}     \\
			WASP-6 & 5450 $\pm$ 100	& 4.6 $\pm$ 0.2	& 1.0 $\pm$ 0.2 & -0.20 $\pm$ 0.09 & \text{\citet{2009AA...501..785G}}     \\
			WASP-17 & 6509 $\pm$ 86	& 4.14 $\pm$ 0.03	& 2.1252	& –0.02 $\pm$ 0.09 & \text{\citet{2012ApJ...757..161T}}    \\
			HAT-P-1 & 5975 $\pm$ 120	& 4.45  $\pm$ 0.15	& 1.295	& +0.13 $\pm$ 0.08 & \text{\citet{2008ApJ...677.1324T}}    \\
			HD 209458 & 6052 $\pm$ 25 & 4.34 $\pm$ 0.028	& 1.50 $\pm$ 0.7 & 0.06 $\pm$ 0.01 & \text{\citet{2016ApJS..225...32B}}                                       \\
			HD 189733 & 5023 $\pm$ 25 & 4.51 $\pm$ 0.028	& 0.8 $\pm$ 0.7 & 0.07 $\pm$ 0.01 &  \text{\citet{2016ApJS..225...32B}}                                         \\
			WASP-19 & 5503 $\pm$ 25	& 4.45 $\pm$ 0.028	& 0.917 $\pm$ 0.7 & 0.22 $\pm$ 0.01 & \text{\citet{2016ApJS..225...32B}}                                      \\
			WASP-12 & 6154 $\pm$ 25 & 4.22 $\pm$ 0.028	& 1.737 $\pm$ 0.7 & 0.22 $\pm$ 0.01 & \text{\citet{2016ApJS..225...32B}}                                    \\
			WASP-43 & 4400 $\pm$ 200 & 4.5 $\pm$ 0.2	& 0.8	& −0.05 $\pm$ 0.17 & \text{\citet{2012AA...542A...4G}}         \\
			WASP-33 & 7471 $\pm$ 63 & 4.35 $\pm$ 0.07 & 2.6218 & 0.1 ([M/H]) & \text{\citet{2010MNRAS.407..507C}}              \\
			\enddata
\end{deluxetable*}

\subsection{Abundance Analysis}
		
To derive local-thermal equilibrium (LTE) abundances for carbon (C), oxygen (O), sodium (Na), sulfur (S), and potassium (K), we follow the procedures described in \citet{2020AJ....159..220S} and \citet{2023ApJ...952...71S}. We measure the equivalent width (EW) of each line manually using the {\it splot} tool in IRAF\footnote{IRAF is distributed by the National Optical Astronomy Observatories, which are operated by the Association of Universities for Research in Astronomy Inc., under a cooperative agreement with the National Science Foundation.}. We then adopt the Kurucz line list, Kurucz stellar atmosphere, and the {\it abfind} task in MOOG (\citealt{2012ascl.soft02009S}) to derive stellar chemical abundances A(X)\footnote{A(X) = 12 + log(N$_X$/N$_H$), where N$_X$ is number of atoms of species X.}. The wavelength in \AA, atomic species, excitation potential (EP) in eV, and log (gf) value for the lines are shown in the headers of Table \ref{tab:eqw} and \ref{tab:individual}. Equivalent width measurements and LTE abundances for each individual line are shown in Table \ref{tab:eqw} and Table \ref{tab:individual} in the \hyperref[appendix]{Appendix}. This traditional abundance derivation technique is not effective for $T_{\rm eff}$ below 4300 K, so we do not provide EW and abundance measurements for K2-18b and GJ3470b.
		
Non-LTE (NLTE) effects can be substantial for elements such as O and S, while being relatively less pronounced for C, Na, and K. Generally, NLTE effects become more pronounced for stars with higher $T_{\rm eff}$ and lower metallicities, as one can also tell from Table 1. We utilize empirical corrections interpolated from \citet{2022AJ....164...87K}, to derive NLTE corrections as listed in Table \ref{tab:NLTE}. In instances where a line's abundance deviates significantly from other lines of the same species, we exclude it as an outlier from the average abundance calculation and mark it with an asterisk in Table \ref{tab:individual}. Adopting the solar abundance from \citet{2009ARAA..47..481A} (also adopted by MOOG), we compute the average [X/H]\footnote{Metallicity [X/H] = A(X)${star}$ -- A(X)$\odot$} in the linear space for C, O, Na, S. There is only a single line available for K so we report [K/H] corresponding to the sole line. It is worth noting that MOOG provides A(X), which we use to normalize the planetary abundance in later discussions, independent of the solar abundance values we chose. The C abundances display $T_{\rm eff}$ dependence, especially for very cool ($<$ 4800 K) or hot ($>$ 7000 K) stars, so we did not report [C/H] for these stars. The final adopted host stellar abundances are shown in Table \ref{tab:abundance}.

Our error measurement process involves several steps. First, we calculate the standard deviation of the mean of [X/H] ($\sigma_{\mu}$) to account for variations in abundance measurements derived from different lines. This computation entails determining the standard deviation of the abundance derived from individual lines of the same species divided by the square root of the number of lines used (e.g. \citealt{2022MNRAS.513.5387S}). Due to the limited number of lines available for some elements, the resulting $\sigma_{\mu}$ can vary widely due to small number statistics. Additionally, we factor in errors propagated from stellar atmospheres. In our study's temperature range, a variation of 100 K in $T_{\rm eff}$ corresponds to an abundance change ranging from 0.01 to 0.06 dex, a change of 0.2 in log {\it g} corresponds to an abundance alteration between 0.003 and 0.06 dex, while a change of 0.2 km s$^{-1}$ in $V_{\rm t}$ results in an abundance shift between 0.01 and 0.05 dex. The final error (except for K) shown in Table \ref{tab:abundance} are co-added in quadrature from $\sigma_{\mu}$ and those propagated from stellar atmospheres. For K where only one line is available that $\sigma_{\mu}$ is underived, we add in quadrature the 1$\sigma$ error from equivalent width measurement with errors propagated from stellar atmospheres.
		
\begin{longrotatetable}
		\begin{splitdeluxetable*}{c|cc|cc|cc|cc|cc|cc|cc|cc|cc|B|c|cc|cc|cc|cc|cc|cc|cccc}
				\label{tab:abundance}
				\tabletypesize{\tiny}
				\tablecaption{[C/H], [O/H], [Na/H], [S/H], [K/H] abundances and literature values}
				\tablehead{
					\colhead{Star Name} & \multicolumn2c{HARPS$^1$} & \multicolumn2c{FEROS$^1$} & \multicolumn2c{HIRES} & \multicolumn2c{HARPS} & \multicolumn2c{FEROS} &  \multicolumn2c{HIRES}  & \multicolumn2c{HARPS} & \multicolumn2c{FEROS}	& \multicolumn2c{HIRES} & Star Name & \multicolumn2c{HARPS} & \multicolumn2c{FEROS}	& \multicolumn2c{HIRES} & \multicolumn2c{HARPS} &  \multicolumn2c{FEROS} & \multicolumn2c{HIRES} &  \multicolumn4c{Literature values$^4$} \\
					\colhead{} & \colhead{[C/H]$^2$} & \colhead{$\sigma^3$} & \colhead{[C/H]} & \colhead{$\sigma$} & \colhead{[C/H]} & \colhead{$\sigma$} & \colhead{[O/H]} & \colhead{$\sigma$} & \colhead{[O/H]} & \colhead{$\sigma$} & \colhead{[O/H]} & \colhead{$\sigma$} & \colhead{[Na/H]} & \colhead{$\sigma$}	& \colhead{[Na/H]} & \colhead{$\sigma$}	& \colhead{[Na/H]} & \colhead{$\sigma$}	& & \colhead{[S/H]} & \colhead{$\sigma$}	& \colhead{[S/H]} & \colhead{$\sigma$}	& \colhead{[S/H]} & \colhead{$\sigma$}	& \colhead{[K/H]} & \colhead{$\sigma$}	& \colhead{[K/H]} & \colhead{$\sigma$}	& \colhead{[K/H]} &\colhead{$\sigma$}	& \colhead{[C/H]} & \colhead{[O/H]}  & \colhead{[Na/H]} & \colhead{[K/H]} 
				}
				\colnumbers
				\startdata
				K2-18	 &   --	  & --	  &  --	   & --	   &       -- &	   --	&	-- & --	&    --	&    --	&    --	&   --	 &	-- & --	&     -- &	   -- &	   --  &	   --	&K2-18	 &	    --	&    --	&      -- &	   -- &	   -- &	   -- &	   -- &	   -- &	   --	&     -- &     -- &	    -- & -- & -- & -- & -- \\                                                                                  
				GJ3470	 &   --	  & --	  &  --	   & --	   &       -- &	   --	&	-- & --	&    --	&    --	&    --	&   --	 &	-- & --	&     -- &	   -- &	   --  &	   --	&GJ3470	 &	    --	&    --	&      -- &	   -- &	   -- &	   -- &	   -- &	-- &	   --	&     -- &     -- &	    -- & -- & -- & -- & -- \\                                                                                  
				HAT-P-26 & 	 --	  & --	  & -0.004	& 0.112 &       -- &	   --	&	-- & --	& 0.064	& 0.174	&    --	&   --	 &	    -- &    --	& -0.059 &	0.118 &	   --  &	   --	&HAT-P-26 &	    --	&    --	&   0.386 &	0.212 &	   -- &	   --	&    -- &    -- & -0.079 &  0.11 &    -- &	    -- & 0.11$^a$ & 0.24$^a$ & 0.02$^a$ & -- \\                                                       
				HAT-P-11 & 	 --	  & --	  &  --	   & --	   &    -- &	 --	&	-- & --	&    --	&    --	& 0.320	& 0.057	 &	    -- &    --	&     -- &	   -- &	0.385  &	0.139	&HAT-P-11 &	    --	&    --	&      --  &   -- &	0.469 &	0.121	&     -- &    -- &     --  &    -- &	-0.078 &    0.13 & -0.168 & 0.10$^b$ & 0.08$^b$ & -- \\                                                    
				WASP-107 &  -- &	-- &	--  & -- &	   -- &	   --	&	-- & --	& 0.281	& 0.105 &    --	&   --	 &	-0.109 & 0.130	& -0.035 &	0.151 &	   --  &	   --	&WASP-107 &	 0.604	& 0.172	&   0.794 &	0.219 &	   -- &	   --	&     -- &    -- & -0.198 &    0.141 &	    -- &    -- & −0.066$^c$ & −0.037$^c$ & +0.161$^c$ & +0.014$^c$ \\                                                                                  
				WASP-127 & -0.340 &	0.081 & -0.246 & 0.082 &	   -- &	   --	&	-- & -- & -0.12	& 0.065	&    --	&   --	 &	-0.270 & 0.086	& -0.315 &	0.095 &	   --  &	   --	&WASP-127 &	-0.365	& 0.126 &  -0.253 &	0.125 &	   -- &	   --	&     -- &    -- & -0.262 &    0.10 &	-- &    -- & -0.12$^a$ & -0.01$^a$	& -0.28$^a$ & -- \\                                                   
				HAT-P-12 & 	 --	  & --	  & --	   & --	   &    -- &	--	&	-- & --	&    --	&    --	& -0.083 & 0.295 &		-- &    --	&     -- &	   -- &	-0.052 &	0.180	&HAT-P-12 &	    --	&    --	&    --	   & --	  & 0.406 &	0.197	&     -- &    -- &     -- &    -- &	-0.385 & 0.12 & -- & -- & -- & -- \\                                                                                  
				WASP-39	 &   --   &	--    &	-0.119	& 0.133 &	   -- &	   --	&	-- & --	& 0.078	& 0.110 &    --	&   --	 &	    -- &    --	& -0.143 &	0.109 &	   --  &	   --	&WASP-39	 &	    --	&    -- &  -0.074 &	0.141 &	   -- &	   --	&     -- &    --  &  0.249 &  0.141  & -- &    -- & -0.04$^a$ & 0.04$^a$ & -0.02$^a$, -0.04$^d$ & -- \\                                      
				WASP-31	 & -0.137 &	0.144 &	   --  &    -- &	   -- &	   --	&	-- & --	&    --	&    --	&    --	&   --	 &	-0.165 & 0.113	&     -- &	   -- &	   --  &	   --	&WASP-31	 &	-0.443	& 0.119 &    --	   & --	  &    -- &	   --	&     -- &    -- &     -- &    -- &	    -- &    -- & -0.08$^e$, -0.12$^b$ & 0.07$^e$, 0.04$^b$ & -0.16$^e$ & -- \\                            
				WASP-96  &     -- &	   -- &	-0.027	& 0.217 &	   -- &	   --	&	-- & --	& 0.283	& 0.206 &    --	&   --	 &	    -- &    --	&  0.136 &	0.205 &	   --  &	   --	&WASP-96  &	    --	&    --	&  0.031 &	0.230 &	   -- &	   --	&     -- &    -- &  0.311 &    0.135 &	    -- &    -- & -- & -- & -- & -- \\                                                                                  
				WASP-6	 & -0.275	& 0.091 &	   --  &    -- &	   -- &	   --	&	-- & --	&    --	&    --	&    --	&   --	 &	-0.165 & 0.109	&     -- &	   -- &	   --  &	   --	&WASP-6	 &	-0.196	& 0.122	&    --	   & --	  &    -- &	   --	& -0.769 &  0.14 &     -- &    -- &	    -- &    -- & & & -0.17$^f$ & -- \\                                                                        
				WASP-17	 & -0.089 &	0.093 &	   --  &    -- &	   -- &	   --	&	-- & --	&    --	&    --	&    --	&   --	 &	-0.011 & 0.159	&     -- &	   -- &	   --  &	   --	&WASP-17	 &	-0.747	& 0.249	&    --	   & --	  &    -- &	   --	& -0.636 &    0.13 &     -- &    -- &	    -- &    -- & -0.44$^e$, -0.48$^b$ & 0.09$^e$, 0.06$^b$ & -0.2$^e$ & -- \\                             
				HAT-P-1	 &     -- &	   -- &	   --  &    -- &	0.042 &	0.182	&	-- & --	&    --	&    --	& 0.206	& 0.089	 &	    -- &    --	&     -- &	   -- &	-0.108 &	0.098	&HAT-P-1	 &	    --	&    --	&    --	   & --	  & 0.213 &	0.113	&     -- &    -- &     -- &    -- &	 0.053 &    0.12 & 0.07$^a$ & 0.21$^a$ &	0.08$^a$ & -- \\                                                  
				HD209458 & -0.095 &	0.045 & -0.079 & 0.050 &	   -- &	   --	&	-- & --	& 0.029	& 0.040	&    --	&   --	 &	-0.147 & 0.050	& -0.156 &	0.043 &	   --  &	   --	&HD209458 &	-0.246	& 0.086 &  -0.107 &	0.086 &	   -- &	   --	&     -- &    -- & -0.141 & 0.10 & -- &    -- & 0.00$^a$, 0.03$^e$, -0.01$^b$ & 0.13$^a$, 0.09$^e$, 0.09$^b$ & -0.03$^a$ , -0.01$^e$ & -- \\ 
				HD189733 &  0.117	& 0.103 & 0.096 &	0.095 &	   -- &	   --	&	-- & --	& 0.122	& 0.044 &    --	&   --	 &	-0.107 & 0.073	& -0.074 &	0.124 &	   --  &	   --	&HD189733 &	 0.327	& 0.096	&  0.302 &	0.164 &	   -- &	   --	&     -- &    -- & -0.102 &  0.11 &	-- &    -- & -0.12$^a$, 0.04$^e$, 0.02$^b$ & -0.21$^a$, 0.07$^e$, -0.01$^b$ & -0.03$^a$, 0.05$^e$ & -- \\ 
				WASP-19	 & 0.121 &	0.049 &	0.052  & 0.070 &	   -- &	   --	&	-- & --	& -0.101 & 0.103	&    --	&   --	 &	 0.175 & 0.074	&  0.114 &	0.040 &	   --  &	   --	&WASP-19	 &	 0.441	& 0.240	&  0.080 &	0.164 &	   -- &	   --	&  0.356 &    0.145 &  0.316 &  0.13 &	    -- &    -- & 0.13$^e$, 0.09$^b$ & 0.18$^e$,	0.15$^b$ & 0.17$^e$ & -- \\                               
				WASP-12	 &     -- &	   -- &	0.049  & 0.067 &	   -- &	   --	&	-- & --	& 0.239	& 0.040 &    --	&   --	 &	    -- &    --	& -0.069 &	0.044 &	   --  &	   --	&WASP-12	 &	    --	&    -- &  -0.055 &	0.079 &	   -- &	   --	&     -- &    -- & -0.294 &  0.11 &	    -- &    -- & 0.09$^e$, 0.07$^b$ & 0.33$^e$, 0.22$^b$ & 0.19$^e$ & -- \\                               
				WASP-43	 &  -- & -- &	   --  &    -- &	   -- &	   --	&	-- & --	&    --	&    --	&    --	&   --	 &	-0.188 & 0.040	&     -- &	   -- &	   --  &	   --	&WASP-43	 &	 0.784	& 0.184	&     -- &    -- &	   -- &    -- &	   -- &    -- &	   --	&     -- &     -- &	    -- & -- & -- & -- & -- \\                                                                              
				WASP-33	 &     -- &	   -- &	   --  &    -- &   -- &	-- & -- & --	&    --	&    --	& 0.532	& 0.432	&	    -- &    --	&     -- &	   -- &	0.215  &	0.246	& WASP-33 &	    --	&    --	&     -- &	   -- &	   -- &	   --	&     -- &    -- &     -- &    -- &	    -- &    -- & -- & -- & -- & -- \\                                                                              
				\enddata
				\tablecomments{1}{For stars with abundance measurements from both HARPS and FEROS spectra, the FEROS abundances are used.}
				\tablecomments{2}{A(X) = 12 + log(N$_X$/N$_H$), where where N$_X$ is number of atoms of species X. [X/H] = A(X)$_{star}$ - A(X)$_{\odot}$, we compute the average [X/H] in the linear space for C, O, Na, S.}
				\tablecomments{3}{For C, O, Na, S, the symbol $\sigma$ represents the co-added error from the standard deviation of the mean of the lines used and the errors propagated from stellar atmospheres. For K, there is only one line available, we add in quadrature the 1$\sigma$ error from equivalent width measurement with errors propagated from stellar atmospheres.}
				\tablecomments{4}{References for stellar abundances and atmosphere: a. \text{W19}; a. \text{\citet{2022RNAAS...6..155P}}; b. \text{\citet{2017arXiv171204944H}}; c. \text{\citet{2023ApJ...949...79H}}; d. \text{\citet{2011AA...531A..40F}}; e. \text{\citet{2016ApJS..225...32B}}; f. \text{\citet{2009AA...501..785G}}.}
			\end{splitdeluxetable*}
\end{longrotatetable}
		
\begin{figure}
	\centering
	\includegraphics[width=0.5\textwidth]{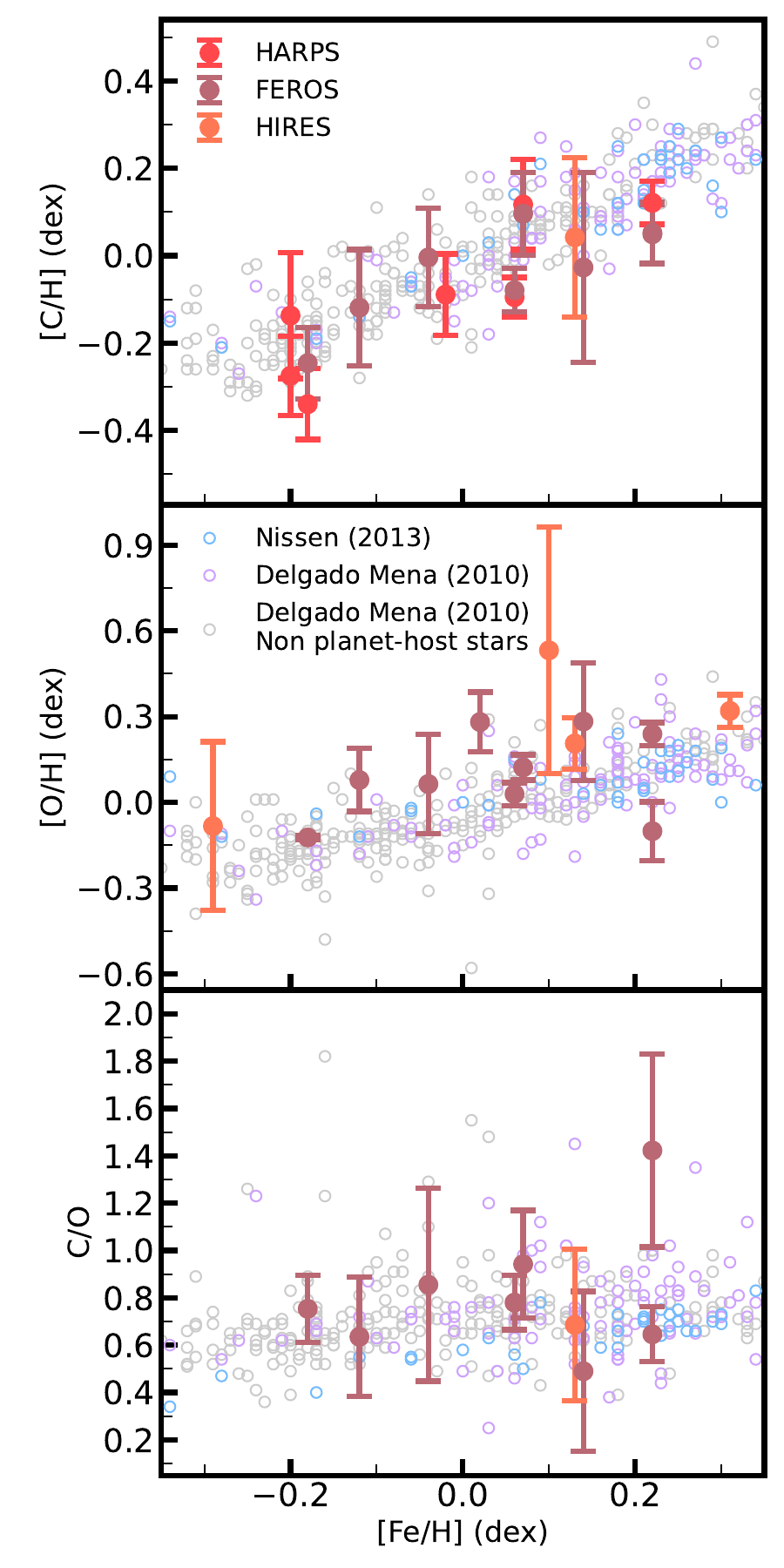}
	\caption{[C/H], [O/H], and C/O ratios plotted against [Fe/H] for our planet-hosting stars using spectra from HARPS, FEROS, and HIRES. The blue open circles show C and O abundances for planet-hosting stars from \citet{2013AA...552A..73N}, while the light purple and gray open circles show planet-hosting and non-planet host stars, respectively, from \citet{2010ApJ...725.2349D}. The background stars in our study exhibit a $T_{\rm eff}$ range of approximately 5000-6500 K, whereas our stars have a much broader $T_{\rm eff}$ range spanning from 3500-7500 K. The error bars are combined from $\sigma_{\mu}$ and error propagated from stellar atmospheres. In the works of \citet{2013AA...552A..73N} and \citet{2010ApJ...725.2349D}, errors for individual stars are not provided. Instead, they estimate typical $\sigma_{\mu}$ to be around 0.1 dex and 0.2 dex for [C/H] and [O/H], respectively.}
	\label{fig:CO}
\end{figure}

When comparing our abundances with those from the literature (the final three columns of Table \ref{tab:abundance}), we note that most of our values align within our error margins, yet some show discrepancies. This underscores the importance of conducting a homogeneous analysis, as systematic errors likely cause inhomogeneities between literature values. In Figure \ref{fig:CO}, we show our final [C/H], [O/H], and C/O ratio against [Fe/H], and compare them to the Galactic sample from \citet{2013AA...552A..73N} and \citet{2010ApJ...725.2349D}. Our C, O abundances, and C/O ratio generally agree with the Galactic sample. However, considering our stars' broader $T_{\rm eff}$ range (3500-7500 K) compared to the Galactic sample (5000-6500 K), some cooler/hotter stars may exhibit more significant deviations in abundances and ratios. Upon closer inspection, our [C/H] values may appear consistently slightly lower, and [O/H] values may be consistently slightly higher, potentially resulting in lower overall C/O ratios. If true, this aligns with findings by \citet{2016ApJ...831...20B}, who reported systematically lower C/O ratios compared to earlier studies. However, these trends may be just attributed to potential visual calibration biases.

\subsection{The Mass-Metallicity Trend for Transiting Exoplanets}
		
\begin{figure*}
	\centering
	\includegraphics[width=1.0\textwidth]{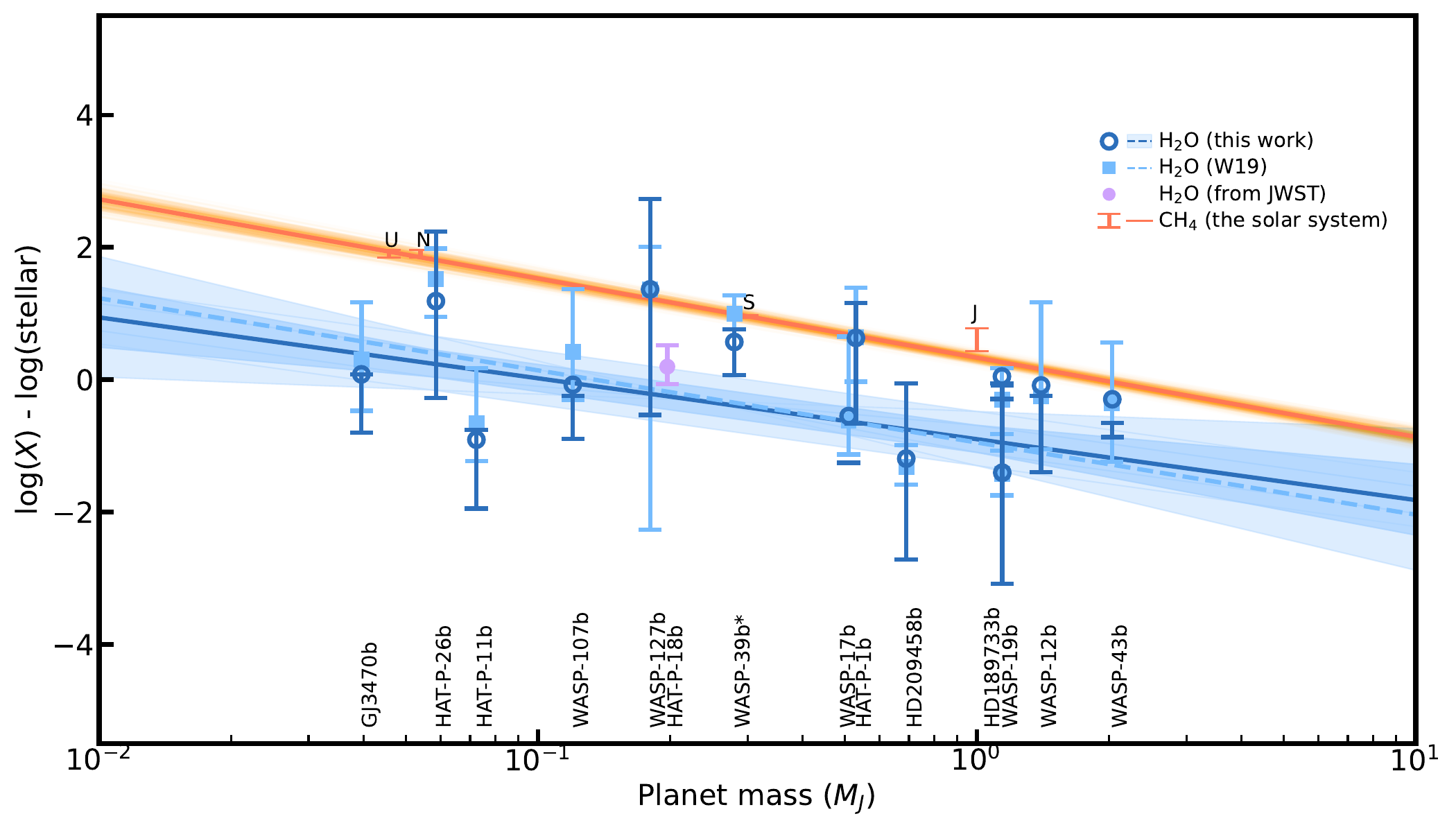}
	\caption{The mass-metallicity correlation among planets, with the x-axis representing planet mass in Jupiter masses ($M_J$) and the y-axis denoting normalized planetary abundance, both are in logarithmic scales. Our normalized H$2$O abundances are denoted by blue open circles, while the solid blue line represents the best-fit linear regression. The shaded blue region delineates the 1$\sigma$ and 2$\sigma$ quantiles. The trend depicted in W19 is shown by the dotted blue line, which exhibits a slightly more significant slope than ours. The linear model in our analysis is expressed as log(H$2$O) = -0.92$^{+0.32}_{-0.33}$log(M/$M_J$) -- 0.90$^{+0.21}_{-0.20}$. A recent detection of H$_2$O from JWST/NIRISS for HAT-P-18b (\citealt{2022ApJ...940L..35F}) is shown in the light purple circle in Figure \ref{fig:trend}.}
	\label{fig:trend}
\end{figure*}

We employ the homogeneous host stellar abundances to normalize the planet atmospheric abundances, in the same way as described in W19. Errors are propagated from both the planet and its host star, with the planetary errors predominantly contributing. If a star had abundance measurements from both the HARPS and FEROS spectra, we used the abundances from FEROS because of the higher SNR. Only planets with chemical detections above the 2$\sigma$ significance were included in our analysis and shown in Figure \ref{fig:trend}. In addition, we exclude K2-18 from the analysis, as \citet{2023ApJ...956L..13M} recently show solid evidence that there is no water detected on K2-18 and the previous claim of water detection was in fact due to methane.

We perform a linear regression analysis for H$_2$O and show the new mass-metallicity relation in Figure \ref{fig:trend} as a blue line. The 1$\sigma$ and 2$\sigma$ confidence intervals are denoted by the blue shaded region. For asymmetric error bars, we use the largest error bar as the value in the $\chi^2$ calculation. The new mass-metallicity relation is slightly less significant than the original relation in W19. The solar CH$_4$ abundances deviate from this population at the 6.4$\sigma$ level. JWST has a higher precision in abundance measurement than that from HST, in this Figure we show the H$_2$O abundance of HAT-P-18b with JWST/NIRISS data (\citealt{2022ApJ...940L..35F}) as an example. This highlights the importance of ensuring homogeneity in host stellar abundance measurement. As the precision in planetary atmospheric measurements improves, discrepancies arising from the systematics in their host stars' abundance can grow notably significant.
		
We compare the Bayes factors between a linear regression model and an independent distribution model for H$_2$O, Na, and K (\citealt{2008arXiv0804.3173R}) to test the robustness of the trend. A Bayes factor greater than 1 ($K$ $>$ 1) indicates that the linear model is preferred over the constant model, while $K$ values between 1 and 10$^{1/2}$ are considered weak evidence. $K$ values between 10$^{1/2}$ and 10 indicate substantial evidence in favor of the linear model, while $K$ values greater than 10 provide strong evidence that the linear model is significantly better. We exclude K2-18, GJ3470, WASP-17, and WASP-43 from our analysis for H$_2$O since their spectra (HARPs) do not cover oxygen lines. Similarly, for Na and K, we only use planets with known host stellar abundance. 

\begin{figure*}
	\centering
	\includegraphics[width=1.0\textwidth]{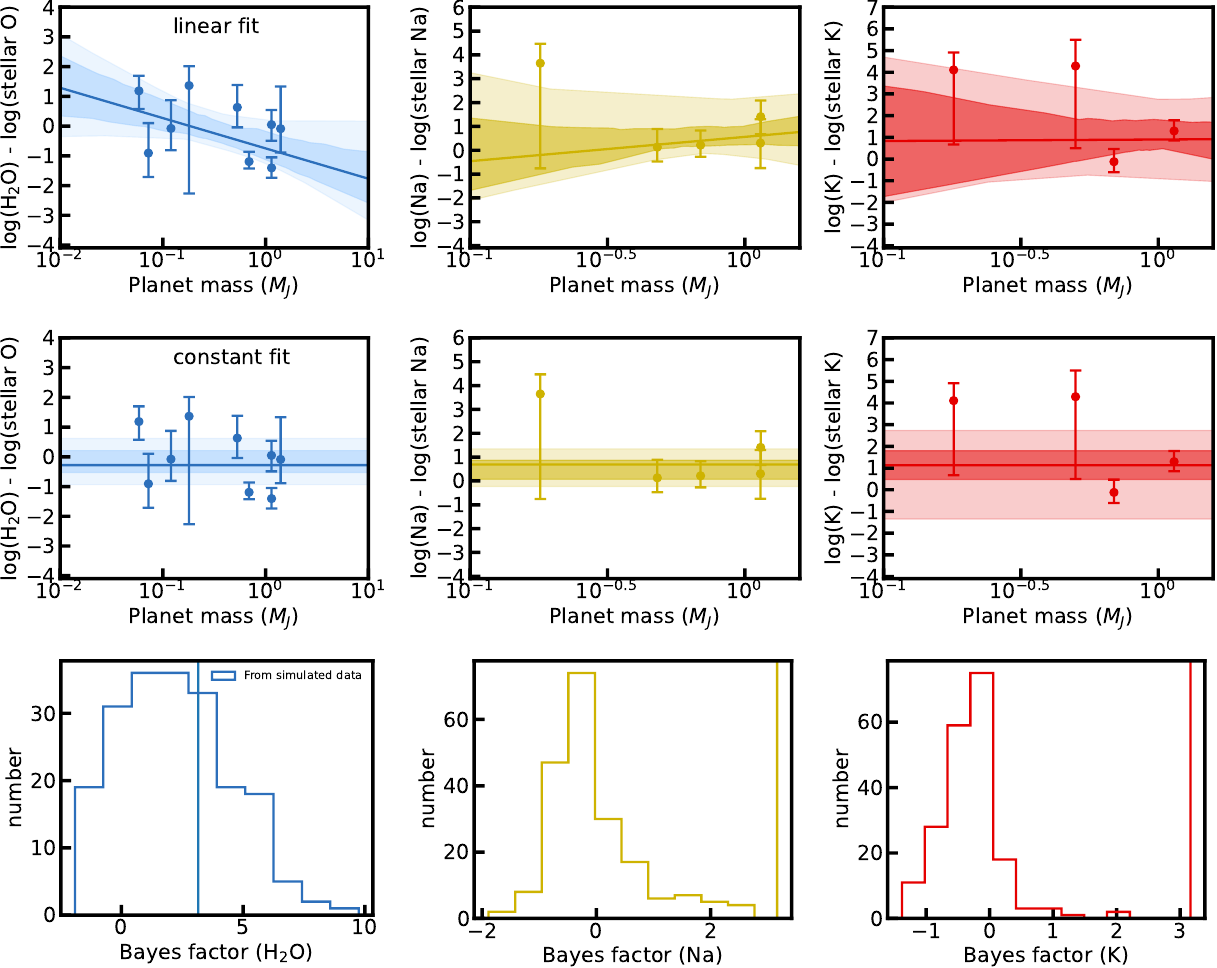}
	\caption{This figure compares the linear regression model (top panel) and the independent distribution model (middle panel) for H$_2$O, Na, and K. The 1$\sigma$ and 2$\sigma$ quantiles are shown in the color-shaded region. The bottom panel displays the Monte Carlo simulation distributions, which were generated by performing 200 tests using the 10,000 data points generated from the linear correlation. We depict the $K = 10^{1/2}$ line, none of the distributions indicate substantial or strong evidence favoring the linear model over the constant model.}
	\label{fig:comb_fit}
\end{figure*}

In Figure \ref{fig:comb_fit}, we show the comparison between the linear regression model (top panel) and the independent distribution model (constant fit, middle panel). For the H$_2$O, Na, and K mass-metallicity trends, the Bayes factors yield values of 1.3, 1.1, and 1.24, respectively. None of these factors suggest a strong or substantial evidence favoring the linear model over the independent distribution model. Since the error in planet mass compared to the error in abundance is relatively small, we neglect the errors in planet mass in our analysis.
		
We subsequently conduct Monte Carlo simulations for a more rigorous examination. Specifically, we randomly select nine data points for H$_2$O, five for Na, and four for K from the 10,000 data points generated using the derived correlation. We repeat this process 200 times and plot the resulting Bayes factor distributions in the histograms shown in the bottom panel of Figure \ref{fig:comb_fit}. The fractions of Bayes factors exceeding 10$^{1/2}$ are consistently low across H$_2$O, Na, and K. Moreover, none of the Bayes factor distributions display a value surpassing 10, refuting strong evidence for a mass-metallicity trend.

The linear mass-$CH_4$ trend (in logarithmic space) in the solar system, supported by four data points, has been used as likely evidence for core accretion. In our study, we conduct a similar Monte Carlo simulation and discover that the distribution of Bayes factors offers substantial evidence with 93\% values exceeding K $>$ 10$^{1/2}$, albeit not strong evidence (32\% have K $>$ 10), in favor of the linear trend between mass and $CH_4$.

Then we conduct similar Monte Carlo simulations but using data generated from the independent distribution model. This allows us to determine how often we might claim a detection by chance for a given threshold. We find false detection rates of 1.5\%, 1.0\%, and 2.0\% for H$_2$O, Na, and K, respectively.
		
If such a correlation truly exists, how many planets and/or how small the individual errors are needed to report a statistically significant detection? To answer this, we perform analysis to determine the minimum number of planets and the required precision of individual measurements needed to confidently detect a correlation ($K$ $>$ 10, strong evidence) between planet mass and metallicity. We perform Monte Carlo simulations using the current error bars and increase the sample size. We find that the Bayes factor distributions are greater than 10 at the 2$\sigma$ confidence interval ($>$95\%) when we increase the sample sizes to 58 for H$_2$O. We do not perform such an analysis for Na and K due to the limitations in our current sample size and the large error bars. It is possible that a trend exists for Na and K but we are not able to detect it. Alternatively, for H$_2$O we simulate using a different error distribution $\sim$ N(0, 1), which is typical for JWST, slightly smaller than the errors in W19, which are all from HST. We find that the abundance of 45 planets and their host stars (H$_2$O and [O/H]) are needed to make the Bayes factors $>$ 10 at the 2$\sigma$ confidence interval.

Furthermore, we examine the relationship between planet atmospheric abundance with other planetary and stellar parameters, including equilibrium temperature ($T_{\rm eq}$), orbital period, eccentricity, semi-major axis, effective temperature ($T_{\rm eff}$) of the host star, planet radius in terms of Jupiter radii ($R_J$), [Fe/H] of the host star, and mass of the host star in solar masses ($M_{\odot}$). Our analysis reveals no significant dependence of H$_2$O, Na, and K abundances on any of these parameters (see Figure~\ref{fig:trend_all} in the \hyperref[appendix]{Appendix}).

With JWST now conducting its initial observations of exoplanet atmospheres, many recent studies have investigated the composition of exoplanet atmospheres (e.g., HAT-P-18b, \citealt{2022ApJ...940L..35F}; WASP-39b, \citealt{2023Natur.614..670F}). We emphasize the importance to derive their host stellar properties including chemical abundances in a homogeneous manner. We call for more and better data to confirm whether a planetary mass-metallicity correlation truly exists, to help distinguish the two planet formation scenarios, including core accretion and disk instability.
		
\section{Summary and Discussion}
		
We obtain HARPS and/or FEROS stellar spectra from the ESO Archive, or reduced HIRES spectra from the KOA, for the host stars of 19 exoplanets from W19. To derive the LTE abundances from equivalent width measurements, we adopt the Kurucz linelist and stellar atmosphere and used MOOG. We then apply empirical non-LTE corrections to report the final [C/H], [O/H], [Na/H], [S/H], and [K/H] abundances. Our final [C/H], [O/H], and C/O ratio are consistent with those of the Galactic sample. Using the updated stellar abundances, we have re-evaluated the mass-metallicity relationship for the 19 transiting planets in W19. Our analysis yield a mass-metallicity trend slightly less significant than the one reported in W19.

In the logarithmic space, we compare the Bayes factors between the linear model and the independent distribution model using only the planets that have corresponding host stellar abundances. None of the Bayes factor distributions of H$_2$O, Na, and K show substantial or strong evidence supporting the preference of linear model over the independent model. Through Monte Carlo simulations using artificially generated data, we determine that to confidently detect a mass-metallicity trend with strong evidence for H$_2$O (Bayes factor $>$ 10 at the 95\% confidence interval), 58 planets with HST precision or 45 planets with JWST precision are needed. A larger quantity of abundance data of higher quality, for both the planets and host stars, is necessary to distinguish between the linear and independent distribution models. The total number of planets from JWST Cycle-1 and Cycle-2 programs is 185, including both transmission or emission spectroscopy and direct imaging. This number has already far exceeded 45. While not every planet may have observable water or carbon carriers, this demonstrates that reaching 45 is not an unattainable goal. Confident detections of any existing mass-metallicity relationship would significantly enhance our understanding of planet formation, and to distinguish between core accretion and disk instability. Future ground-based high-resolution spectroscopy could have better constraints on Na and K (e.g. \citealt{2022AJ....164..173C}), and this HST sample is the best sample to start with as combing low-res with high-res provides the best constraints on alkali lines for the atmospheres of these hot Jupiters.

We are optimistic about the potential of JWST in advancing our understanding of mass-metallicity trends. Preliminary measurements from the JWST already indicate improved precision and a broader range of molecular species, including carbon-bearing compounds, which can be used to test mass-metallicity relationships. By combining high-quality data with robust statistical methods, we anticipate significant progress in unraveling the mysteries of planetary system formation and evolution beyond our own.

\clearpage
\section*{acknowledgements}
			
We acknowledge the science research grants from the China Manned Space Project with NO. CMS-CSST-2023. Q.S. thanks support from the Shuimu Tsinghua Scholar Program.
			
This work is based on data obtained from the ESO Science Archive Facility (\citealt{https://doi.org/10.18727/archive/33, https://doi.org/10.18727/archive/24}). This research has made use of the Keck Observatory Archive (KOA) (\citealt{Tran2014}), which is operated by the W. M. Keck Observatory and the NASA Exoplanet Science Institute (NExScI), under contract with the National Aeronautics and Space Administration. Some of the data presented herein were obtained at the W. M. Keck Observatory, which is operated as a scientific partnership among the California Institute of Technology, the University of California and the National Aeronautics and Space Administration. The Observatory was made possible by the generous financial support of the W. M. Keck Foundation. The authors wish to recognize and acknowledge the very significant cultural role and reverence that the summit of Maunakea has always had within the indigenous Hawaiian community.  We are most fortunate to have the opportunity to conduct observations from this mountain.
			
		
\bibliography{sun22_planet}{}
\bibliographystyle{aasjournal}
		
\appendix  \label{appendix}

\begin{longrotatetable}
\begin{deluxetable*}{c|ccccccccccccccccc}
		\label{tab:eqw}
		\tablecaption{Equivalent width of C, O, Na, S, K lines}
		\tabletypesize{\tiny}
		\tablehead{
					\colhead{\tiny{Wavelength}} & \colhead{5052.167} & \colhead{5380.337} & \colhead{6587.610} & \colhead{7111.470} & \colhead{7113.179} & \colhead{7771.944}	& \colhead{7774.166} & \colhead{7775.388} & \colhead{4751.822} & \colhead{5148.838} & \colhead{6154.225} & \colhead{6160.747} & \colhead{6046.000} & \colhead{6052.656} & \colhead{6743.540} & \colhead{6757.153} & \colhead{7698.974} \\
					\colhead{\tiny{Species}} &\colhead{6.0} & \colhead{6.0} & \colhead{6.0} & \colhead{6.0} & \colhead{6.0} & \colhead{8.0}	& \colhead{8.0} & \colhead{8.0} & \colhead{11.0} & \colhead{11.0} & \colhead{11.0} & \colhead{11.0} &  \colhead{16.0} &  \colhead{16.0} &  \colhead{16.0} &  \colhead{16.0} & \colhead{19.0} \\
					\colhead{\tiny{EP}} &\colhead{7.685} & \colhead{7.685} & \colhead{8.537} & \colhead{8.64} & \colhead{8.647} & \colhead{9.146}	& \colhead{9.146} & \colhead{9.146} & \colhead{2.1044} & \colhead{2.1023} & \colhead{2.1023} & \colhead{2.1044} &  \colhead{7.87} &  \colhead{7.87} &  \colhead{7.87} &  \colhead{7.87} & \colhead{0.000} \\
					\colhead{\tiny{log (GF)}} &\colhead{-1.24} & \colhead{-1.57} & \colhead{-1.05} & \colhead{-1.07} & \colhead{-0.76} & \colhead{0.37}	& \colhead{0.22} & \colhead{0.0} & \colhead{-2.078} & \colhead{-2.044} & \colhead{-1.547} & \colhead{-1.246} &  \colhead{-0.15} &  \colhead{-0.40} &  \colhead{-0.60} &  \colhead{-0.15} & \colhead{-0.168} 
				}
				\startdata
				\multicolumn{14}{c}{HARPS} \\
				\hline
				K2-18 & 7 & 1 & 4 & --&--&--&--&--&--&--&--&--& -- & -- & -- & -- & --\\
				GJ3470	&  12 & 9 & 3 &--&--& -- & -- & -- &--&--&--&--& -- & -- & -- & -- & -- \\
				WASP 107 & 18 & 16 & 4 & --&--&--&--&--& 42 & 64 & 111 & 120 & 2 & 2 & 2 & 4 & -- \\
				WASP 127 & 30 & 13 & 10 & --&--&--&--&--& 11 & 7 & 26 & 43 & 18 & 7 & 9 & 10 & --\\
				WASP 31	& 35 & 33 & 16 &  --&--&--&--&--& 7 & 9 & 16 & 28 & 20 & 15 & 6 & 25 & --\\
				WASP 6 & 15 & 7 & 4 & -- & -- & -- & -- & -- & 20 & 15 & 36 & 57 & 7 & 5 & 5 & 7 &132 \\
				WASP 17 & 60 &40& 11 & -- & -- & -- & -- & -- & 5 & 11 & 30 & 23 & 3 & 10 & 6 & 5 & 103 \\ 
				HD209458 & 40 & 25 & 16 & --&--&--&--&--& 7 & 8 & 27 & 48 & 13 & 15 & 7 & 16 & -- \\
				HD189733 & 15 & 10 & -- & --&--&--&--&--& 19 & 35 & 64 & 84 & 13 & 5 & 5 & 3 & -- \\
				WASP 19	& 34 & 19 & 10 & -- & -- & -- & -- & -- & 35 & 23 & 55 & 80 & 16 & 8 &20 & 35 & 215 \\
				WASP 43 & 10 & 4 & 4 & -- & -- & -- & -- & -- & 50 & 60 & 137 & 120 & 3 & 4 & 3 & 2 &--\\
				\hline
				\multicolumn{14}{c}{FEROS} \\
				\hline
				HAT-P-26 & 10 & 8 & 3 & 4 & 7 & 26 & 30 &12 & 18 & 35 & 69 & 82 & 7 & 6 & 10 & 6 & 246\\
				WASP-107 &  14 & 18 & 4 & 8 & 13 & 7 & 5 & 4 & 47 & -- & 118 & 138 & 2 & 3 & 3 & 6 & 491 \\
				WASP-127 & 31 & 15 & 10 & 13 & 15 & 80 & 67 & 52 & 10 & 6 & 27 & 37 & 17 & 16 & -- & 14 & 152 \\
				WASP-39 & 17 & 10 & 5 & 3 & 14 & 44 & 40 & 33 &15 & 15 & 47 & 66 & 12 & 5 & 2 & 14 &211 \\
				WASP-96	& 22 & 14 & 9 & 13 & 8 & 65 & 63 & 40 &22&22&70&76& 15 & 6 & 7 & 22 & 190\\
				HD 209458 & 40 & 24 & 17 & 18 & 20 & 100 & 91 & 74 & 7 & 8 & 27 & 46 & 18 & 17 & -- & 22 & 150 \\
				HD 189733 & 12 & 9 & 3 & 3 & 10 & 26 & 21 & 14 & 21 & 38 & 67 & 83 & 10 & -- & 6 & 6 & 244 \\ 
				WASP-19 & 25 & -- & 10 & --& 17 & 35 & 40 & 28 & 46 & 25 & 58 & 76 & 11 & -- & 10 & 15 & 208 \\
				WASP-12 & 58 & 37 & 31 & 16 & 30 & 124 & 109 & 90 & 8 & 11 & 34 & 48 & 27 & 22 & 14 & 39 & 135 \\
				\hline
				\multicolumn{14}{c}{HIRES} \\
				\hline
				HAT-P-11 & 7 & 19 & 1 & 11 & 23 & 21 &18 & 12 & 54 & 92 & 120 & 121 & 5 & 3 & 3 & 8 & 313 \\
				HAT-P-12 & 11 & 27 & 5 & 4 & 21 & 4 & 3 & 7 & 43 & 80 & 75 & 89 & 3 & 3 & 3 & -- & 293 \\
				HAT-P-1 & 45 & 9 & 24 & 14 & 36 & 107 & 98 & 73 & 31 & 9 & 33 & 46 & 22 & 18 & 17 & 28 & 157 \\
				WASP-33 & 23 & 25 & 1 & 2 & 1 & 206 & 300 & 135 & 1 & 5 & 23 & 40 & 23 \\
				\enddata
				\tablecomments{1}{}
			\end{deluxetable*}
		\end{longrotatetable}
		
\begin{longrotatetable}
		\begin{deluxetable*}{c|ccccccccccccccccc}
				\label{tab:individual}
				\tablecaption{Abundances$^1$ of individual C, O, Na, S, K lines}
				\tiny
				\tablehead{
					\colhead{\tiny{Wavelength}} & \colhead{5052.167} & \colhead{5380.337} & \colhead{6587.610} & \colhead{7111.470} & \colhead{7113.179} & \colhead{7771.944}	& \colhead{7774.166} & \colhead{7775.388} & \colhead{4751.822} & \colhead{5148.838} & \colhead{6154.225} & \colhead{6160.747} & \colhead{6046.000} & \colhead{6052.656} & \colhead{6743.540} & \colhead{6757.153} & \colhead{7698.974} \\
					\colhead{\tiny{Species}} &\colhead{6.0} & \colhead{6.0} & \colhead{6.0} & \colhead{6.0} & \colhead{6.0} & \colhead{8.0}	& \colhead{8.0} & \colhead{8.0} & \colhead{11.0} & \colhead{11.0} & \colhead{11.0} & \colhead{11.0} &  \colhead{16.0} &  \colhead{16.0} &  \colhead{16.0} &  \colhead{16.0} & \colhead{19.0} \\
					\colhead{\tiny{EP}} &\colhead{7.685} & \colhead{7.685} & \colhead{8.537} & \colhead{8.64} & \colhead{8.647} & \colhead{9.146}	& \colhead{9.146} & \colhead{9.146} & \colhead{2.1044} & \colhead{2.1023} & \colhead{2.1023} & \colhead{2.1044} &  \colhead{7.87} &  \colhead{7.87} &  \colhead{7.87} &  \colhead{7.87} & \colhead{0.000} \\
					\colhead{\tiny{log (GF)}} &\colhead{-1.24} & \colhead{-1.57} & \colhead{-1.05} & \colhead{-1.07} & \colhead{-0.76} & \colhead{0.37}	& \colhead{0.22} & \colhead{0.0} & \colhead{-2.078} & \colhead{-2.044} & \colhead{-1.547} & \colhead{-1.246} &  \colhead{-0.15} &  \colhead{-0.40} &  \colhead{-0.60} &  \colhead{-0.15} & \colhead{-0.168} 
				}
				\startdata
				\multicolumn{14}{c}{HARPS} \\
				\hline
				K2-18	&	-- & -- &	-- &	-- & --	&     -- &     -- &     -- & -- & -- & -- & -- & -- & -- & -- & -- & --    \\
				GJ3470	& -- &	-- & -- &	-- & --	& -- & -- & -- & -- & -- & -- & -- & -- & -- & -- & -- & --  \\
				WASP-107  &	-- & -- & -- &	-- & --	&     -- &     -- &     -- & 6.043 & 6.301 & 6.360 & 6.149 & 7.366 & 7.615 & 7.915 & 7.815 & --  \\
				WASP-127  &	8.162 &	8.016 &	8.138 &	-- & --	&     -- &     -- &     -- & 6.193 & 5.928 & 6.057 & 6.062 & 6.998 & 6.746 & 7.093 & 6.696 & --  \\
				WASP-31	  &	8.158 &	8.463 &	8.225 &	-- & --	&     -- &     -- &     -- & 6.227 & 6.296 & 6.062 & 6.067 & 6.939 & 7.029 & 6.784 & 7.095 & --    \\
				WASP-6 & 8.190 &	8.145 &	8.158 &	-- & --	&  -- &  -- &  -- & 6.323 & 6.127 & 6.090 & 6.118 & 6.924 & 7.009 & 7.245 & 6.961 & -- \\
				WASP-17	  &	8.355 &	8.387 &	7.829* &	-- & --	&  -- &  -- &  -- & 6.158 & 6.475 & 6.481 & 6.031 & 5.855 & 6.665 & 6.629 & 6.092 & -- \\
				HD 209458 &	8.349 &	8.379 &	8.336 &	-- & --	&     -- &     -- &     -- & 6.114 & 6.124 & 6.220 & 6.290 & 6.793 & 7.119 & 6.949 & 6.932 & --    \\
				HD 189733 &	8.481 &	8.622 &	 -- &	-- & --	&     -- &     -- &     -- & 6.042 & 6.333 & 6.265 & 6.241 & 7.580 & 7.311 & 7.580 & 6.872 & --    \\
				WASP-19	  &	8.612	& 8.576	& 8.484 &	-- & --	&     -- &     -- &     -- & 6.674 & 6.38  & 6.437 & 6.510 & 7.268 & 7.131 & 7.925 & 7.878 & 5.496 \\
				WASP-43	 &	-- & -- & -- &	-- & --	&     -- &     -- &     -- & 6.125 & 6.219 & 6.154 & 6.101 & 7.613 & 8.006 & 8.164 & 7.512 & --  \\
				\hline
				\multicolumn{14}{c}{FEROS} \\
				\hline
				HAT-P-26  &	8.194	& 8.448	& 8.297	& 8.556	& 8.531 &	8.777	& 9.039	& 8.604 &	6.054 &	6.372 &	6.374 &	6.250 & 7.177 & 7.347 & 7.885 & 7.156 &	5.101 \\
				WASP-107  &	-- & -- & -- & -- &	-- &	8.989 &	8.930 &	9.018 &	6.114 &	   -- &	6.431 &	6.314 & 7.366 & 7.813 & 8.118 & 8.035 &	4.942 \\
				WASP-127  &	8.184 &	8.094 &	8.138 &	8.382 &	8.158 &	8.803 &	8.761 &	8.735 &	6.147 &	5.857 &	6.078 &	5.963 & 6.965 & 7.180 & -- & 6.880 & 5.168 \\
				WASP-39	  &	8.233 &	8.295	& 8.238	& 8.107	& 8.577 &	8.811 &	8.877 &	8.937 &	6.154 &	6.103 &	6.260 &	6.250 & 7.185 & 6.982 & 6.786 & 7.324 &	5.309 \\
				WASP-96	  &	8.251	& 8.339	& 8.384	& 8.692	& 8.129 &	9.05	& 9.166	& 8.944 &	6.400 &	6.349 &	6.651 &	6.438 & 7.186 & 6.940 & 7.262 & 7.487 &	5.361 \\
				HD 209458 &	8.349 &	8.353 &	8.371 &	8.496 &	8.253 &	8.969 &	9.006 &	8.992 &	6.114 &	6.124 &	6.220 &	6.259 & 6.972 & 7.188 & -- & 7.123 & 5.389 \\
				HD 189733 &	8.346	& 8.563	& 8.349	& 8.474	& 8.786 &	8.875	& 8.859	& 8.788 &	6.094 &	6.384 &	6.308 & 6.228 & 7.429 & -- & 7.676 & 7.224 & 5.068 \\
				WASP-19	  &	8.388 &	 -- & 8.484 & -- &	8.582 &	8.505 &	8.771 &	8.691 &	6.855* & 6.427 &	6.483 &	6.451 & 7.053 & -- & 7.504 & 7.290 & 5.456 \\
				WASP-12	 &	8.544 &	8.536 &	8.629 &	8.326 &	8.392 &	9.131	& 9.108	& 9.086 &	6.222 &	6.321 &	6.397 &	6.325 & 7.125 & 7.249 & 7.219 & 7.416 &	5.256 \\
				\hline
				\multicolumn{14}{c}{HIRES} \\
				\hline
				HAT-P-11 & -- & -- & -- & -- & -- & 9.190 &  9.221 &	9.144 &	6.491 &	6.949 &	6.786 &	6.495 & 7.401 & 7.391 & 7.681 & 7.760 &	5.032 \\
				HAT-P-12 & -- & -- & -- & -- & -- & 8.255 &  8.243 &	8.962 &	6.149 &	6.634 &	6.063 &	5.955 & 7.208 & 7.457 & 7.747 & -- &	4.715 \\
				HAT-P-1	 & 8.511 & 7.873 & 8.647 & 8.424 & 8.683 & 9.141 &  9.186 &	9.074 &	6.834* & 6.147 &	6.307 &	6.228 & 7.151 & 7.282 & 7.486 & 7.348 &	5.413 \\
				WASP-33	 & -- & -- & -- & -- & -- & 9.204 & 10.013 &	8.817 &	5.823 & 6.490 & 6.718 & 6.730 & 5.302 & 5.551 & 5.757 & 5.305 &	-- \\
				\enddata
				\tablecomments{1}{A(X) = 12 + log(N$_X$/N$_H$), where where N$_X$ is number of atoms of species X.}
				\tablecomments{*}{We exclude the line when computing abundance average in Table \ref{tab:abundance}, either because they deviate too much from the abundance of the other lines, or their EW measurements are unreliable.}
			\end{deluxetable*}
\end{longrotatetable}
		
\begin{figure*}
	\centering
	\includegraphics[width=1.0\textwidth]{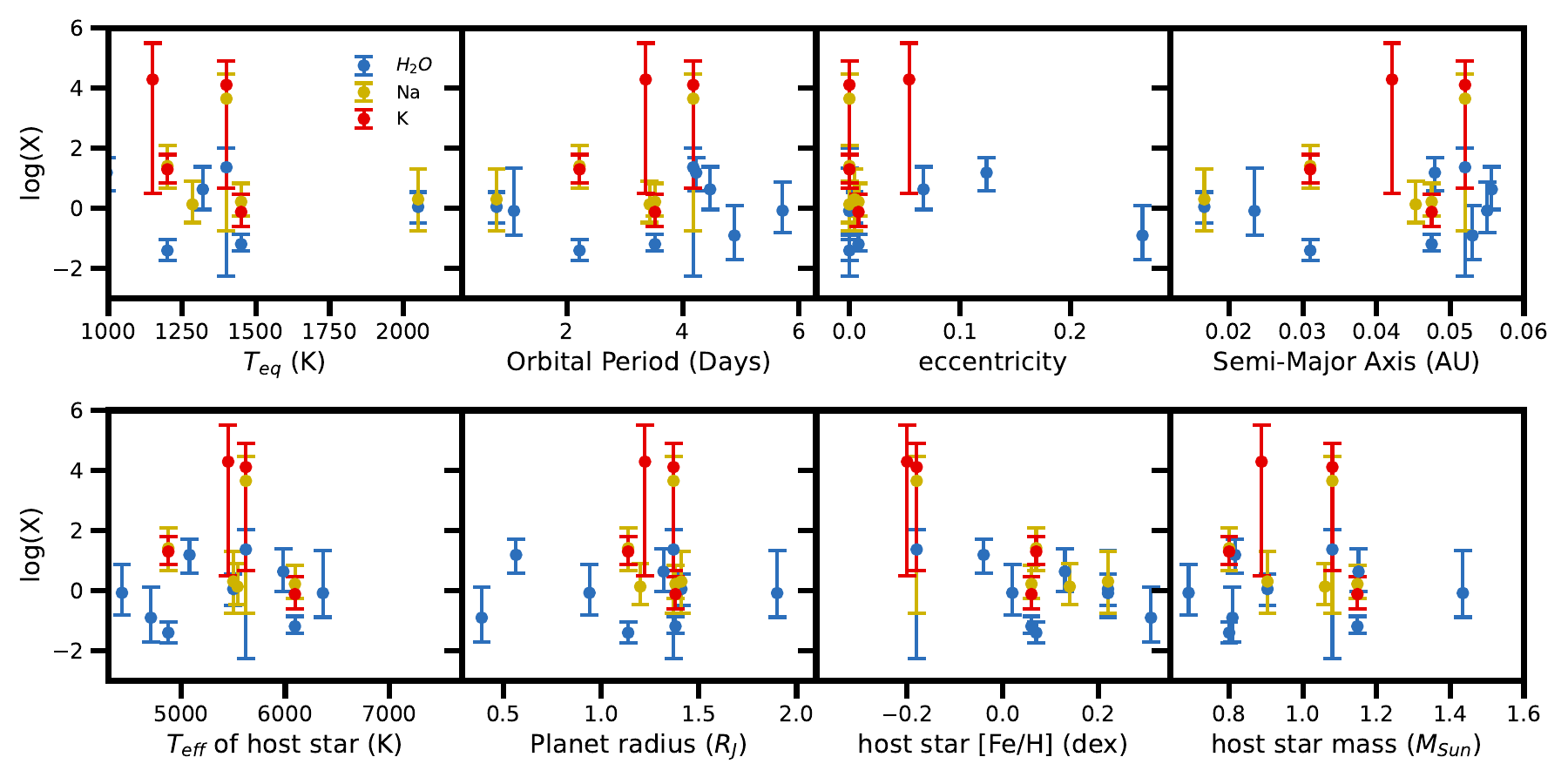}
	\caption{The normalized planetary atmosphere H$_2$O, Na, K abundance vs other planetary/stellar parameters.}
	\label{fig:trend_all}
\end{figure*}
		
\end{document}